\documentclass[10pt,final,journal]{IEEEtran}

\usepackage{amsthm}
\usepackage{amssymb}
\usepackage{amsmath}

\newtheorem{theorem}{Theorem}
\newtheorem{lemma}{Lemma}
\newtheorem{corollary}{Corollary}

\newcommand{\Aut}{\ensuremath{\mathrm{Aut}}}

\newcommand{\wt}{\ensuremath{\mathrm{wt}}}
\newcommand{\F}{\ensuremath{\mathbb{F}}}

\begin{document}

\title{On Optimal Binary One-Error-Correcting Codes
of Lengths $2^m-4$ and $2^m-3$}

\author{Denis S. Krotov, Patric R. J. \"Osterg{\aa}rd,
and Olli Pottonen
\thanks{The work of the first author was supported by the Federal Target Program
``Scientific and Educational Personnel of Innovation Russia''
for 2009--2013 (government contract No.\ 02.740.11.0429) and the
Russian Foundation for Basic Research under
Grant No.\ 10-01-00424.
The work of the second author was financed by the Academy of Finland
under Grants No.\ 130142 and 132122.
The work of the third author was supported by the Academy of Finland
Grant No.\ 128823, Helsinki Institute of Information Technology HIIT project
``Algorithmic Systems'', and the Finnish Cultural Foundation.
}
\thanks{D. Krotov is with the Sobolev Institute of Mathematics,
and with the Mechanics and Mathematics Department,
Novosibirsk State University, 630090 Novosibirsk, Russia.}
\thanks{P. R. J. {\"O}sterg{\aa}rd is with the
Department of Communications and Networking, Aalto University
School of Electrical Engineering, P.O.\ Box 13000, 00076 Aalto, Finland.
He was also with Lehrstuhl Mathematik II, Universit{\"a}t Bayreuth,
95440 Bayreuth, Germany.}
\thanks{O. Pottonen was with the Department of Information and Computer Science,
Aalto University School of Science, P.O.\ Box 15400, 00076 Aalto, Finland.
He is now with the Departament de Llenguatges i Sistemes Inform\`atics,
Universitat Polit\`ecnica de Catalunya,
Jordi Girona 1--3, 08034 Barcelona, Spain.
}
}
\date{}

\maketitle

\begin{abstract}
Best and Brouwer [\emph{Discrete Math.} {\bf 17} (1977), 235--245]
proved that triply-shortened and doubly-shortened
binary Hamming codes (which have length $2^m-4$ and $2^m-3$,
respectively) are optimal. Properties of such codes are here studied,
determining among other things
parameters of certain subcodes. A utilization of these properties makes
a computer-aided classification of the optimal binary
one-error-correcting codes
of lengths $12$ and $13$ possible; there are $237610$ and $117823$
such codes, respectively (with $27375$ and $17513$ inequivalent
extensions). This completes the classification of optimal
binary one-error-correcting codes for all lengths up to $15$.
Some properties of the classified codes are further investigated.
Finally, it is proved that for any $m \geq 4$,
there are optimal binary one-error-correcting codes of length
$2^m-4$ and $2^m-3$ that cannot be lengthened to perfect codes
of length $2^m-1$.
\end{abstract}


\section{Introduction}

\IEEEPARstart{A}{binary code}
of length $n$ is a set $C \subseteq \F_2^n$, where
$\F_2 = \{0,1\}$ is the field of order $2$.
The (Hamming) \emph{distance} between
elements ${\bf c},{\bf c'} \in \F_2^n$, called \emph{words}
(or \emph{codewords} when they belong to a code),
is the number of coordinates
in which they differ and is denoted by $d({\bf c},{\bf c'})$.
The \emph{minimum distance} of a code is the smallest pairwise
distance among distinct codewords:
\[
d(C) = \min\{d({\bf c},{\bf c'}) : {\bf c}, {\bf c'} \in C,\ {\bf c} \neq {\bf c'}\}.
\]
The (Hamming) \emph{weight} $\wt({\bf c})$ of a word
${\bf c} \in \F_2^n$ is the number of nonzero coordinates.

A binary code of length $n$, size $M$, and minimum distance
$d$ is said to be an $(n,M,d)$ code. Since a code with minimum
distance $d$ is able to correct up to $\lfloor (d-1)/2 \rfloor$
errors, such a code is said to be
$\lfloor (d-1)/2 \rfloor$-error-correcting. 
If every word in the ambient space is
at distance at most
$\lfloor (d-1)/2 \rfloor$ from some codeword of a
$\lfloor (d-1)/2 \rfloor$-error-correcting code, then the
code is called \emph{perfect}.

The maximum size of a
binary code of length $n$ and minimum distance $d$ is
denoted by $A(n,d)$; the corresponding codes are said to be
\emph{optimal}. For binary codes there
is a direct connection between optimal error-correcting codes
with odd and even minimum distance:
\begin{equation}
\label{eq:34}
A(n+1,2d) = A(n,2d-1).
\end{equation}
One gets from the odd case to the even case by \emph{extending}
the code with a parity bit, and from the even case to the
odd case by removing an arbitrary coordinate, called
\emph{puncturing}. Other transformations of codes include
\emph{shortening}, where a coordinate is deleted and all
codewords but those with a given value in the deleted
coordinate are removed, and \emph{lengthening} which is the
reverse operation
of shortening. See \cite{MS77} for
the basic theory of error-correcting codes.

When studying optimal error-correcting
codes---or suboptimal for that sake---it is reasonable
to restrict the study to codes that are essentially
different in the following sense.
Two binary codes are said to be \emph{equivalent} if the codewords
of one of the codes can be mapped onto those of the other by
the addition of a vector followed by a permutation of the coordinates.
Such a mapping from a code onto itself is an \emph{automorphism}
of the code; the set of all automorphisms of a code $C$ forms
the \emph{automorphism group} of $C$, denoted by $\Aut(C)$.

A code with only even-weight codewords is said to be \emph{even}.
Codes equivalent to even codes are of central importance in the
current work; these codes have only even-weight codewords or only
odd-weight codewords, and they are characterized by the fact that
the distance between any two codewords is even. We therefore call such
codes \emph{even-distance codes} (not to be confused
with codes that have even minimum distance).

Hamming codes are perfect (and thereby optimal)
one-error-correcting codes:
\[
A(2^m-1,3) = 2^{2^m-m-1}.
\]
Best and Brouwer \cite{BB77} showed
that by shortening Hamming codes one, two, or three times, one
still gets optimal codes:
\begin{equation}
\label{short}
A(2^m-1-i,3) = 2^{2^m-m-1-i},\quad 0 \leq i \leq 3.
\end{equation}

For all but the very smallest parameters, there are many inequivalent
codes with the parameters in (\ref{short}). In general, a complete
characterization or classification of such codes does not seem
feasible, but the classification problem can be addressed for small
parameters and general properties of these codes can be studied.
For example, the issue whether codes with these parameters can be
lengthened to perfect codes has attracted some interest in the
literature
\cite{B99,EV98,OP10,K10a}.
For $i=1$, every code
(\ref{short}) can be lengthened to a perfect code and this can be
done in a unique way up to equivalence \cite{B99}. Consequently, codes with
such parameters are in a direct relationship to the perfect codes,
so our main interest is in the codes with $i=2$ and $i=3$.

One aim of the current work is to study properties of codes with
the parameters of doubly-shortened and triply-shortened perfect
binary one-error-correcting codes. This study is started
in Section~\ref{sect:prop} by considering certain properties
of subcodes, which can be utilized in a computer-aided
classification of optimal binary one-error-correcting codes of
length $12$ and $13$, considered in Section~\ref{sect:all}. It turns out that
the number of equivalence classes of $(12,256,3)$ and $(13,512,3)$ codes is
$237610$ and $117823$, respectively. Some
central properties of the classified codes are analyzed in
Section~\ref{sect:res}. Finally, infinite families of optimal
one-error-correcting codes of length $2^m-4$ and $2^m-3$
that cannot be lengthened to perfect one-error-correcting
codes of length $2^m-1$ are presented in Section~\ref{sect:lengthen}.
A preliminary version of some of the results in this work can
be found in \cite{K10a}.

As only binary codes are considered
in the current work, the word binary is omitted in the sequel.

\section{Properties of Subcodes}

\label{sect:prop}

Some properties related to subcodes of the codes under study are
conveniently investigated in the framework of orthogonal arrays.
An $\mbox{OA}_\lambda (t,k,q)$
\emph{orthogonal array} of index $\lambda$, strength $t$,
degree $k$, and order $q$ is a $k \times N$ array
with entries from $\{0,1,\ldots ,q-1\}$ and the property that
every $t \times 1$ column vector appears exactly $\lambda$ times
in every $t \times N$ subarray; necessarily $N = \lambda q^t$.

The \emph{distance distribution} $(A_0,A_1,\ldots A_n)$ of an
$(n,M,d)$ code $C$ is defined by
\[
A_i = \frac{1}{M}|\{({\bf c},{\bf c'}) : {\bf c},{\bf c'} \in C,\ d({\bf c},{\bf c'})=i\}|.
\]
We will need the following theorem by Delsarte~\cite{D73}; for
more information about the MacWilliams transform, see also
\mbox{\cite[Chapter 5]{MS77}}.

\begin{theorem}
\label{thm:delsarte}
An array is an orthogonal array of strength\/ $t$
if and only if the MacWilliams transform of the distance distribution
of the code formed by the columns of the array has entries\/
$A'_0 = 1$, $A'_1 = A'_2 = \cdots = A'_t = 0$.
\end{theorem}

We are now ready to prove a central result, essentially
following the arguments of \cite[Theorem 6.1]{BB77} (where,
however, the case $d=3$ rather than $d=4$ is considered).

\begin{theorem}
\label{thm:main}
Every\/ $(2^m-3,2^{2^m-m-4},4)$ code is an even-distance code
and forms an\/ $\mbox{\rm OA}_{\lambda}(t,n,2)$ with\/
$t=2^{m-1}-4$, $n=2^m-3$, and\/
$\lambda = 2^{2^{m-1}-m}$.
\end{theorem}

\begin{IEEEproof}
We first show that an even-distance $(n=2^m-3,M=2^{2^m-m-4},4)$ code $C$
forms an orthogonal array with the given parameters. Let $A_i$
be the distance distribution of $C$, and let $A'_i$ be the MacWilliams
transform of $A_i$, that is,
\begin{eqnarray}
\label{eq:B=A}
MA'_k & = & \sum_{i=0}^n A_i K_k(i),\nonumber\\
\label{eq:B=A'}
2^nA_k & = & M\sum_{i=0}^n A'_i K_k(i),
\end{eqnarray}
where
\[
K_k(i) = \sum_{j=0}^k(-1)^j \binom{i}{j}\binom{n-i}{k-j}
\]
is a Krawtchouk polynomial. It is well known that $A'_0 = 1$ and
$A'_i \geq 0$ for $1 \leq i \leq n$
\cite{D73}.

As $C$ is an even-distance code, $A_i = 0$ for odd $i$, and,
since $K_{n-k}(i) = (-1)^{i} K_{k}(i)$, we have
\begin{equation}
\label{eq:mirror}
A'_{k} = A'_{n-k}.
\end{equation}

Let $\alpha(i) = (n-3) K_0(i)+2K_2(i)+2K_{n-1}(i)$.
Direct calculations now show that
\begin{equation}
\label{eq:defal}
\alpha(i) = (n-2i-2+(-1)^i)(n-2i+2+(-1)^i).
\end{equation}
From (\ref{eq:defal}) and $n = 2^m-3 \equiv 1 \pmod 4$ we derive
\begin{eqnarray}
\alpha((n-3)/2) = \alpha((n-1)/2) = \nonumber \\
\label{eq:4cases}
\alpha((n+1)/2)= \alpha((n+3)/2) = 0,
\end{eqnarray}

\noindent
and $\alpha(i)>0$ for any other integer $i$. We have
$A_0=1$, $A_{n-1} \leq 1$, and, since $C$ has
minimum distance $4$, $A_2=0$. Utilizing
(\ref{eq:mirror}), we then get

\begin{eqnarray}
\label{eq:al}
2\alpha(0)A'_0 & = & \alpha(0)A'_0 + \alpha(n)A'_n
\leq \sum_i \alpha(i) A'_i\nonumber \\
& = & \frac{2^n((n-3)A_0+2A_2+2A_{n-1})}{M}\\
& = & \frac{2^n(n-3+2A_{n-1})}{M} \leq \frac{2^n(n-1)}{M}\nonumber
\end{eqnarray}

\noindent
and thereby
\[
M \leq \frac{2^n(n-1)}{2\alpha(0)A'_0} =
\frac{2^n(n-1)}{2(n-1)(n+3)} = \frac{2^{n-1}}{n+3}.
\]

We know that in fact $M = 2^{n-1}/(n+3)$, so we have equalities in
(\ref{eq:al}). This implies that
$\alpha(0)A'_0 + \alpha(n)A'_n = \sum_i \alpha(i) A'_i$, that is,
$\alpha(i)A'_i = 0$ for $1 \leq i \leq n-1$. By
(\ref{eq:4cases})
and the comment thereafter, it follows that
$A'_i = 0$ for $1 \leq i \leq (n-5)/2$ (and $(n+5)/2 \leq i \leq n-1$).
Application of Theorem~\ref{thm:delsarte} shows that we have an
orthogonal array with the given parameters.

To show that any $(2^m-3,2^{2^m-m-4},4)$ code is indeed an
even-distance code, we assume that there is a code $C$ which is not, to
later arrive at a contradiction. The code $C$ can be partitioned
into sets of even-weight and odd-weight codewords, denoted by
$C_{\mathrm{even}}$ and $C_{\mathrm{odd}}$, respectively. That is,
$C = C_{\mathrm{even}} \cup C_{\mathrm{odd}}$, with $|C_{\mathrm{even}}|\geq 1$
and $|C_{\mathrm{odd}}|\geq 1$. For any codewords,
${\bf c} \in C_{\mathrm{even}}$, ${\bf c'} \in C_{\mathrm{odd}}$,
we have $d({\bf c},{\bf c'}) \geq 5$ (as the distance is odd and greater
than 4). Let
\[
C_i = C_{\mathrm{even}} \cup (C_{\mathrm{odd}}+{\bf e}_i),
\]
where ${\bf e}_i$ is the weight-one vector with the 1 in coordinate
$i$. We now know that $C_i$ is an even-distance $(2^m-3,2^{2^m-m-4},4)$ code
for any $1 \leq i \leq n$.

We next prove that $C_{\mathrm{odd}}$ is an orthogonal array with
the same strength $t$ (see the early part of the proof)
as the $n$ different even-distance codes $C_i$.
The proof that the same holds for $C_{\mathrm{even}}$ is analogous.
W.l.o.g., it suffices to consider the last $t$ coordinates and
two $t$-tuples ${\bf t}_1,{\bf t}_2$ that differ only in one
(we choose the last) coordinate---induction then shows that this
holds for any pairs---and show that these two
$t$-tuples occur in equally many codewords of $C_{\mathrm{odd}}$.

We denote the set of words in a code $C$ that have value ${\bf d}$ in
the last $t$ coordinates by $C({\bf d})$. Then
\[
|C_{\mathrm{odd}}({\bf t}_1)| = |(C_{\mathrm{odd}}+{\bf e}_1)({\bf t}_1)| =
| C_1({\bf t}_1) | - | C_{\mathrm{even}}({\bf t}_1)|,
\]
\[
|C_{\mathrm{odd}}({\bf t}_2)| = |(C_{\mathrm{odd}}+{\bf e}_n)({\bf t}_1)| =
| C_n({\bf t}_1) | - | C_{\mathrm{even}}({\bf t}_1)|.
\]
Since $C_1$ and $C_n$ both form orthogonal arrays with strength $t$,
$| C_1({\bf t}_1) | = | C_n({\bf t}_1) |$, and it follows that
$|C_{\mathrm{odd}}({\bf t}_1)| = |C_{\mathrm{odd}}({\bf t}_2)|$.

As $C_{\mathrm{odd}}$ is an even-distance code that forms an orthogonal array
with strength $t=2^{m-1}-4$, we can now reuse the calculations in the
beginning of this proof to determine a lower bound on the size of
$C_{\mathrm{odd}}$. Namely, we now have $\alpha(i) A'_i = 0$ except for
$i = 0$ and $i=n$, and can carry out calculations closely related
to (\ref{eq:al}):
\begin{eqnarray}
2\alpha(0)A'_0 & = & \alpha(0)A'_0 + \alpha(n)A'_n
= \sum_i \alpha(i) A'_i\nonumber \\
& = & \frac{2^n((n-3)A_0+2A_2+2A_{n-1})}{|C_{\mathrm{odd}}|} \nonumber \\
& = & \frac{2^n(n-3+2A_{n-1})}{|C_{\mathrm{odd}}|}
      \geq \frac{2^n(n-3)}{|C_{\mathrm{odd}}|},\nonumber
\end{eqnarray}
\noindent
so
\[
|C_{\mathrm{odd}}| \geq \frac{2^n(n-3)}{2\alpha(0)A'_0} =
\frac{2^{n-1}(n-3)}{(n-1)(n+3)} = |C|\frac{n-3}{n-1}.
\]
But similarly one gets $|C_{\mathrm{even}}| \geq |C|(n-3)/(n-1)$, and thereby
$|C| = |C_{\mathrm{even}}|+|C_{\mathrm{odd}}| = |C|2(n-3)/(n-1) > |C|$ when
$n > 5$, a contradiction.
\end{IEEEproof}

\begin{corollary}
\label{cor:weightdist}
A\/ $(2^m-3, 2^{2^m-m-4}, 4)$ code has a unique distance distribution.
\end{corollary}
\begin{IEEEproof}
It suffices to prove that the MacWilliams transform of the distance distribution
is unique. By the proof of Theorem~\ref{thm:main},
for a $(2^m-3, 2^{2^m-m-4}, 4)$ code we have $A'_k = 0$ for every $k$
except for $A'_0 = A'_n = 1$ and the unknown values
$A'_{(n-1)/2} = A'_{(n+1)/2}$ and
$A'_{(n-3)/2} = A'_{(n+3)/2}$.
Equation~(\ref{eq:B=A'}) with $k=0, 2$ gives a pair of equations
which determines the unknown values.
\end{IEEEproof}

Consequently, the remark at the end of \cite{BB77} about the
distance distribution of certain codes not being unique applies only to
triply-shortened perfect codes and not to triply-shortened extended
perfect codes.

\begin{corollary}
\label{cor:eqeven}
Every\/ $(2^m-i,2^{2^m-m-1-i},4)$ code with\/ $0 \leq i \leq 3$
is an even-distance code.
\end{corollary}

\begin{IEEEproof}
From a code with the given parameters that is not an
even-distance code, one can get a subcode for which the same holds.
This can be done by shortening in a coordinate where two codewords
that are at odd mutual distance have the same value. This is not
possible by Theorem~\ref{thm:main}.
\end{IEEEproof}

The distance-$k$ graph of a code is a graph with one vertex for
each codeword and edges between vertices whose corresponding
codewords are at mutual distance $k$.

\begin{corollary}
Every\/ $(2^m-1-i,2^{2^m-m-1-i},3)$ code with\/ $0 \leq i \leq 3$
has a connected distance-$3$ graph.
\end{corollary}

\begin{IEEEproof}
If the distance-3 graph of an $(n,M,3)$ code is
not connected, then there are more than one way of extending
the code to an $(n+1,M,4)$ code; cf.\ \cite[p.~230]{KO06}.
In particular, it can then be extended to a code that is
not an even-distance code. This is not possible by
Corollary~\ref{cor:eqeven}.
\end{IEEEproof}

\begin{corollary}
\label{cor:shorten}
Shortening a\/ $(2^m-3,2^{2^m-m-4},4)$ code\/
$t$ times with\/ $t \leq 2^{m-1}-4$ gives a\/
$(2^m-3-t,2^{2^m-m-4-t},4)$ code that is
an even-distance code.
\end{corollary}

In particular, with $m=4$ and $t=4$, we always
get a $(9,16,4)$ subcode after shortening a
$(13,256,4)$ code four times.

However, not all $(2^m-3-t,2^{2^m-m-4-t},4)$ codes
with $t \leq 2^{m-1}-4$ are subcodes of some
$(2^m-3,2^{2^m-m-4},4)$ code. We shall now
strengthen the necessary condition in
Corollary~\ref{cor:shorten} for a code to
be a subcode of a $(2^m-3,2^{2^m-m-4},4)$ code.
Since the result is of interest specifically
for the classification in Section~\ref{sect:all},
for clarity it is presented only for subcodes
of $(13,256,4)$ codes.
For the general case, similar conditions can alternatively be
obtained using results by Vasil'eva \cite{V04}
and connections between $(2^m-4,2^{2^m-m-4},3)$ codes and
\mbox{1-perfect} codes of length
$2^m-1$ \cite[Corollary~4]{K10b}.

\begin{theorem}
\label{thm:LC}
Let\/ $C$ be obtained from a\/ $(13,256,4)$ code by
shortening\/ $t$ times, $0 \leq t \leq 4$, and let\/
$N_w$ denote the number of codewords of weight\/ $w$ in\/ $C$.
If\/ $C$ is an even code, then\/ $(5-t)N_0+N_2 \geq 5-t$,
and if\/ $C$ is a code with only odd-weight codewords,
then\/ $(5-t)N_{1}+N_3\leq (t^2-11t+44)/2$.
\end{theorem}

\begin{IEEEproof}
Without loss of generality, we assume that shortening is
carried out by extracting codewords with $0$s in
$t$ given coordinates (after which the $t$ coordinates
are deleted).

We first consider the case $t=0$ given an even
$(13,256,4)$ code. Consider all $\binom{13}{4}$
subcodes obtained by looking at all different sets of $4$ coordinates
and shortening with respect to 0s in these coordinates.
By Corollary~\ref{cor:shorten}, every such subcode has cardinality $16$,
so the sum of their cardinalities is
$\binom{13}{4}\cdot 16 = 11440$.
In this sum, every codeword (in the original code) of weight $0$ is
considered $\binom{13}{4}=715$ times; similarly for
each codeword of weight $2$, $4$, $6$, and $8$, we get the counts
$330$, $126$, $35$, and $5$, respectively.

After repeating these calculations with respect to shortenings in
$3$, $2$, $1$, and $0$ coordinates, we arrive at the following system of
equations:
$$\left[\begin{array}{r@{\ }r@{\ }r@{\ }r@{\ }r@{\ }r@{\ }r}
715\ & 330\ & 126\ & 35\ &  5\ & 0\ & 0\\
286\ & 165\ &  84\ & 35\ & 10\ & 1\ & 0\\
 78\ &  55\ &  36\ & 21\ & 10\ & 3\ & 0\\
 13\ &  11\ &   9\ &  7\ &  5\ & 3\ & 1\\
  1\ &   1\ &   1\ &  1\ &  1\ & 1\ & 1\\
\end{array}\right]
\left[\begin{array}{l}N_0\\N_2\\N_4\\N_6\\N_8\\N_{10}\\N_{12}\\\end{array}\right]
 =
\left[\begin{array}{r}
11440\\
9152\\
4992\\1664\\256
\end{array}\right].
$$
When these equations are combined with the coefficients
$8/128$, $-36/128$, $94/128$, $-187/128$, $315/128$, and with the
coefficients
$8/128$, $-52/128$, $190/128$, $-515/128$, $1155/128$, one gets the
equations $5N_0 + N_2 + N_{12} = 6$ and $N_0+N_{10}+5N_{12} = 22$,
respectively. Since $N_{12}\leq 1$
and $N_0 \geq 0$, we get $5N_0 + N_2 \geq 5$ and $N_{10}+5N_{12} \leq 22$.
From the latter inequality, we get $5N_1 + N_3 \leq 22$
for odd-weight codes after adding
the all-one
word to all codewords.
This completes the proof for $t=0$.

The inequality $5N_0 + N_2 \geq 5$ means that we have
either $N_0 = 1$ or $N_2 \geq 5$ (or both).
In the former case, we will have one codeword of
weight $0$ after any shortening. In the latter case, on the other hand,
the codewords of weight $2$ must have disjoint supports,
so at most $t$ of them are lost when shortening $t$
times. It follows that $(5-t)N_0 + N_2 \geq 5-t$
after shortening $t$ times. This proves the first part of the
theorem.

For the second part of the theorem, we use induction and
let $C$ be a code obtained by shortening an even $(13,256,4)$ code
$t-1$ times. Moreover, let $C = 0C_0 \cup 1C_1$, so $C_0$ and $C_1$
are obtained after shortening the $(13,256,4)$ code $t$ times; $C_0$
is obviously even and $C_1$ has only odd-weight codewords.
We also define
the code $C' = 1C_0 \cup 0C_1$ (which is obviously equivalent to $C$).

The weight distributions of the codes $C$, $C'$, $C_0$, and $C_1$ are
denoted by $N_w$, $N'_w$, $N^{0}_w$, and $N^{1}_w$, respectively, so
$N_w = N^{0}_w + N^{1}_{w-1}$ and $N'_w = N^{0}_{w-1} + N^{1}_w$.
From
\[
(5-t+1)N_0+N_2 \geq 5-t+1
\]
and
\[
(5-(t-1)) N'_1 + N'_3\leq ((t-1)^2-11(t-1)+44)/2,
\]
we now obtain
\begin{eqnarray*}
\lefteqn{(5-t)N^{1}_1+N^{1}_3}\\
&=&((5-(t-1))N^{1}_1+(5-(t-1))N^{0}_0+N^{1}_3+N^{0}_2)\\
&\mbox{}&  -((5-(t-1))N^{0}_0+N^{0}_2+N^{1}_1)\\
&=& ((5-(t-1))N'_1 + N'_3)-((5-(t-1))N_0+N_2)\\
&\leq& ((t-1)^2-11(t-1)+44)/2-(5-(t-1))\\
&=& (t^2-11t+44)/2.
\end{eqnarray*}
This completes the proof.
\end{IEEEproof}

It could be possible to sharpen Theorem~\ref{thm:LC}, but, as we shall
later see, it fulfills our needs in the current study.

\section{Classification of One-Error-Correcting Codes}

\label{sect:all}
Before describing the classification approach used in the current work,
we give a short review of some old related classification results.

\subsection{Survey of Old Results}

A survey of classification results for optimal error-correcting codes
can be found in \cite[Section~7.1.4]{KO06}, where catalogues of
optimal codes can also be obtained in electronic form. In the current
study, we consider optimal codes with $d=3$---that is, optimal
one-error-correcting codes---and $d=4$. Zaremba \cite{Z52} proved that the code attaining
$A(7,3) = 16$ is unique (up to equivalence) and so is therefore its extension;
it is not difficult
to show that all optimal codes with shorter lengths are also unique.
Baicheva and Kolev \cite{BK98} proved that there are $5$ equivalence
classes of codes attaining
$A(8,3) = 20$, and these have $3$ extensions. Litsyn and Vardy \cite{LV94}
proved uniqueness of the code attaining $A(9,3)=40$ and its extension.
The second author of this paper together with Baicheva and Kolev
classified the codes attaining $A(10,3)=72$ and $A(11,3)=144$;
there are $562$ equivalence classes (with $96$ extensions) and $7398$
equivalences classes (with $1041$ extensions) of
such codes, respectively.

Knowing the sizes of the optimal one-error-correcting codes up to length 11,
one in fact knows the sizes of such codes up to length 15 by~(\ref{short}).

The perfect codes attaining $A(15,3) = 2048$
were classified
by the second and the third author \cite{OP09}; the number of equivalence
classes of such codes is $5983$, with $2165$ extensions.
Using a result by Blackmore \cite{B99},
this classification can be used to get the number of equivalence classes of
codes attaining $A(14,3) = 1024$, which is $38408$; these have $5983$ extensions.
All these results still leave the
classification problem open for lengths $12$ and $13$. It is known \cite{OP10}
that not all such codes can be obtained by shortening codes of length $14$ or $15$.

\subsection{Classification Approach}

The general idea underlying the current work is to classify codes
in an iterative manner by utilizing the fact that an $(n,M,d)$ code
has an $(n-1,M',d)$ subcode with $M' \geq M/2$. This idea---with various
variations---has been used earlier in \cite{OBK99} and elsewhere.
However, it is easy to argue why it is not feasible to classify the
$(12,256,3)$ and $(13,512,3)$ codes directly in such a manner.

A classification of the $(12,256,3)$ and $(13,512,3)$ codes
via a classification of the $(11,M',3)$ codes
with $M' \geq 128$ would lead to a prohibitive number of codes of length
11. To see this, it suffices to obtain a rough bound on the number
of equivalence classes of $(11,128,3)$ codes. Every $(11,144,3)$ optimal code has
$\binom{144}{128}$ different subsets of 128 codewords, and any such set
of words can be equivalent to at most $2^{11}11!$ sets in total.
Therefore, there are at least
\[
\frac{\binom{144}{128}}{2^{11}11!}\approx 8.4 \cdot 10^9
\]
equivalence classes of $(11,128,3)$ codes. Similar (rough) bounds
can be obtained for the number of $(11,M,3)$ codes with
$129 \leq M \leq 144$.

So far in this section, we have considered the case $d=3$. Of course,
by (\ref{eq:34}), we might as well consider the case $d=4$. In fact,
we shall do so in the sequel, to get a smaller number of equivalence
classes of subcodes in each stage.

To make the classification feasible, we shall make use of
Corollary~\ref{cor:shorten}, which shows that
not only do all $(12,M,4)$ subcodes
of the $(13,256,4)$ and $(14,512,4)$ codes have $M=128$, but we
have the much stronger result that all $(9,M,4)$ subcodes
of the $(13,256,4)$ and $(14,512,4)$ codes have size $M=16$ and are
even-distance codes. Moreover,
the number of subcodes to be considered can be reduced considerably
by Theorem~\ref{thm:LC}.

All in all, by Corollary~\ref{cor:shorten}
the $(13,256,4)$ and $(14,512,4)$ codes can be obtained as follows:
\begin{equation}
\label{eq:seq}
\begin{array}{l}
(9,16,4) \rightarrow (10,32,4) \rightarrow (11,64,4) \rightarrow\\
(12,128,4) \rightarrow (13,256,4) \rightarrow (14,512,4).
\end{array}
\end{equation}
The even-distance $(9,16,4)$ codes are classified iteratively
from smaller codes,
without any assumptions on the sizes of subcodes.

As described in \cite[Section~7.1.1]{KO06}, lengthening is carried out
by using a clique algorithm. For each set of parameters in the
sequence (\ref{eq:seq}),
the number of codes is further reduced by isomorph rejection and by
discarding codes that do not fulfill Corollary~\ref{cor:shorten} and
Theorem~\ref{thm:LC}. Details regarding the implementation of some of these
parts will be discussed next.

\subsection{Implementation and Results}
\label{sect:result}

Before presenting the results of the computations, we shall consider
some details regarding the implementation of various parts of the
algorithm.

The method of lengthening codes by finding cliques in a certain
\emph{compatibility graph}---consisting of one vertex for each
(even) word that can be added and with edges between vertices whose
corresponding words are at mutual distance at least $d$---is
well known, cf.\ \cite[Section~7.1.1]{KO06}. However, we are here
facing the challenge of finding rather large cliques---up to size
256, in the last step of (\ref{eq:seq}). This clique search can
be sped up as follows in the last three steps of (\ref{eq:seq}),
again relying on the theoretical results.

Consider the step of lengthening an $(n,2^{n-5},4)$ code with
$11 \leq n \leq 13$, by including
a coordinate with 0s for these codewords and adding
codewords of length $n+1$ with 1s in the new (say, first)
coordinate. The candidates for the new codewords can be
partitioned into $2^{n-10}$ sets $S_i$ depending on the values in the
first $n-9$ coordinates (recall that the value in the first coordinate
is 1 for all of these). Let $G_i$ be the subgraph of the original
compatibility graph induced by the vertices corresponding to
the codewords in $S_i$.
We now construct a new graph $G$ with one vertex for all cliques
of size 32 in $G_i$ for any $i$, and with edges between vertices whenever the
corresponding codes pairwise fulfill the minimum distance criterion.
The cliques of size $2^{n-10}$ in $G$
give the desired codes. The program Cliquer \cite{NO03} was used in this
work to solve clique instances.

\emph{Isomorph rejection}, that is, detecting and removing copies of
equivalent codes, is carried out via a transformation into a
graph \cite{OBK99} and using the graph isomorphism program \emph{nauty}
\cite{M90}. The graph considered has two vertices for each coordinate,
one for each value of the coordinate. The program \emph{nauty} can be
asked to give a canonical labeling of the vertices; we use the idea of
canonical
augmentation \cite{M98} and require that the vertex corresponding to
the new coordinate and the value given to the old codewords have the
smallest label. (See \cite{O10} for an analogous approach for constant
weight codes.) Codes that pass this test must still be compared
with the other codes obtained from the same subcode.

For the first few sets of parameters in (\ref{eq:seq}), \emph{nauty}
processes the graphs in a sufficiently fast manner. However, the larger the
codes, the greater is the need for enhancing such a direct approach, cf.\
\cite{OP09}. In the current work, an invariant was used that is
based on sets of four codewords with the same value in all but six coordinates,
where they form the structure \{000000,111100,110011,001111\} \cite{OP09,P00}.

The search starts from the $343566$ equivalence classes of
even-distance $(9,16,4)$ codes, which in turn were classified
iteratively from smaller codes.
In Table~\ref{tab:search}, the number of equivalence classes of codes after
each lengthening and application of the necessary conditions is shown.

\begin{table}[htbp]
\caption{Number of intermediate (even-distance) codes}
\label{tab:search}
\begin{center}
\begin{tabular}{lrr}\hline
$(n,M,d)$ & \# \\\hline
(9,16,4)  & 25170\\
(10,32,4) & 24819\\
(11,64,4) & 31899\\
(12,128,4) & 37667\\
(13,256,4) & 27375\\
(14,512,4) & 17513\\\hline
\end{tabular}
\end{center}
\end{table}

Table \ref{tab:search} shows that there are $27375$
equivalence classes of $(13,256,4)$ codes as well as
$17513$ equivalence classes of $(14,512,4)$ codes.
Puncturing the codes in all possible
ways and carrying out further isomorph rejection
reveals that there are $237610$ equivalence classes of $(12,256,3)$ codes and
$117823$ equivalence classes of $(13,512,3)$ codes.
A total of less than one month of CPU-time using one core of a 2.8-GHz
personal computer was needed for the whole search.

Before presenting the main properties of the classified codes,
we shall briefly discuss validation of these computer-aided results.

\subsection{Validation of Classification}

Data from the classification steps can be used to validate
the results by using a double-counting argument. More
specifically, the total number of even-distance $(n,2^{n-5},4)$ codes
(that is, labeled codes disregarding equivalence) with
$10 \leq n \leq 14$ can be
counted in two ways. This is a well-known technique,
see \mbox{\cite[Chapter~10]{KO06}} and \cite{O10}.

The orbit-stabilizer theorem gives the number of
labeled even-distance $(n,2^{n-5},4)$ codes as
\begin{equation}
\label{eq:sum1}
\sum_{C \in \mathcal{C}} \frac{2^nn!}{|\Aut(C)|},
\end{equation}
where $\mathcal{C}$ is a set with one code from
each equivalence class of such codes.

Let $\mathcal{C'}$ be a set of
representatives from all equivalence
classes of even-distance $(n-1,2^{n-6},4)$ codes and $N_{C}$ the number of
final codes (before isomorph rejection) that are obtained
in the computer search starting from the code $C$. Then the
total number of labeled codes can also be obtained as
\begin{equation}
\label{eq:sum2}
\sum_{C \in \mathcal{C'}} \frac{2^{n-1}(n-1)!N_{C}}{|\Aut(C)|},
\end{equation}
and it can be checked whether $(\ref{eq:sum1}) = (\ref{eq:sum2})$.

For the classification leading up to $(9,16,4)$ codes,
a modified scheme analogous to the that in \cite{O10}
was utilized.

The utilization of Corollary~\ref{cor:shorten}
and Theorem~\ref{thm:LC} in the three steps from $(9,16,4)$
to $(12,128,4)$ implies that not all even-distance
$(n,2^{n-5},4)$ codes
are classified for $10 \leq n \leq 12$. A more extensive
modification of the counting argument, apparently requiring
a modification of the classification scheme as well, would be
necessary to handle these instances; this was not considered
in the current work. In any case, the double-counting argument
gave the desired result for the final two steps, the
classification of $(13,256,4)$ and $(14,512,4)$ codes.

\section{Properties of the Classified Codes}

\label{sect:res}

In Tables \ref{tab:aut1} to \ref{tab:aut4}, the orders of the automorphism
groups of the classified codes are shown.

\begin{table}[htbp]
\begin{center}
\caption{Automorphisms of $(12, 256, 3)$ codes}
\label{tab:aut1}
\begin{tabular}{rrrrrrrr}\hline
$|\Aut(C)|$ & \# & $|\Aut(C)|$ & \# & $|\Aut(C)|$ & \# \\ \hline
1 & 14179 & 64 & 8511 & 2048 & 39 \\
2 & 45267 & 96 & 90 & 3072 & 3 \\
3 & 41 & 128 & 3114 & 4096 & 9 \\
4 & 66449 & 192 & 55 & 6144 & 4 \\
6 & 137 & 256 & 1247 & 8192 & 1 \\
8 & 44529 & 384 & 39 & 12288 & 4 \\
12 & 159 & 512 & 403 & 16384 & 1 \\
16 & 32193 & 768 & 35 & 24576 & 1 \\
24 & 89 & 1024 & 82 & 73728 & 1 \\
32 & 20813 & 1152 & 1 & 147456 & 1 \\
48 & 98 & 1536 & 15 &   &   \\
\hline
\end{tabular}
\end{center}
\end{table}

\begin{table}[htbp]
\begin{center}
\caption{Automorphisms of $(13, 256, 4)$ codes}
\label{tab:aut2}
\begin{tabular}{rrrrrr}\hline
$|\Aut(C)|$ & \# & $|\Aut(C)|$ & \# & $|\Aut(C)|$ & \# \\ \hline
1 & 841 & 64 & 2041 & 3072 & 4 \\
2 & 2781 & 96 & 37 & 4096 & 7 \\
3 & 24 & 128 & 818 & 4608 & 1 \\
4 & 5507 & 192 & 37 & 6144 & 2 \\
6 & 35 & 256 & 395 & 8192 & 1 \\
8 & 5034 & 384 & 19 & 12288 & 2 \\
12 & 39 & 512 & 161 & 16384 & 1 \\
16 & 5352 & 768 & 18 & 24576 & 1 \\
24 & 52 & 1024 & 38 & 73728 & 1 \\
32 & 4043 & 1536 & 17 & 147456 & 1 \\
48 & 50 & 2048 & 15 &   &   \\
\hline
\end{tabular}
\end{center}
\end{table}

\begin{table}[htbp]
\begin{center}
\caption{Automorphisms of $(13, 512, 3)$ codes}
\label{tab:aut3}
\begin{tabular}{rrrrrrrr}\hline
$|\Aut(C)|$ & \# & $|\Aut(C)|$ & \# & $|\Aut(C)|$ & \# \\ \hline
1 & 782 & 64 & 15534 & 3072 & 15 \\
2 & 4464 & 96 & 48 & 4096 & 59 \\
3 & 55 & 128 & 6988 & 6144 & 5 \\
4 & 11412 & 192 & 51 & 8192 & 13 \\
6 & 71 & 256 & 3245 & 12288 & 3 \\
8 & 19902 & 384 & 16 & 16384 & 7 \\
12 & 37 & 512 & 1391 & 24576 & 1 \\
16 & 27406 & 768 & 19 & 32768 & 1 \\
24 & 54 & 1024 & 475 & 49152 & 1 \\
32 & 25506 & 1536 & 26 & 98304 & 1 \\
48 & 73 & 2048 & 162 &   &   \\
\hline
\end{tabular}
\end{center}
\end{table}

\begin{table}[htbp]
\begin{center}
\caption{Automorphisms of $(14, 512, 4)$ codes}
\label{tab:aut4}
\begin{tabular}{rrrrrrrr}\hline
$|\Aut(C)|$ & \# & $|\Aut(C)|$ & \# & $|\Aut(C)|$ & \# \\ \hline
1 & 23 & 96 & 25 & 3072 & 19 \\
2 & 187 & 128 & 2300 & 4096 & 72 \\
3 & 8 & 192 & 51 & 6144 & 8 \\
4 & 599 & 256 & 1429 & 8192 & 23 \\
6 & 31 & 336 & 5 & 12288 & 10 \\
8 & 1167 & 384 & 37 & 16384 & 3 \\
12 & 43 & 512 & 713 & 21504 & 1 \\
16 & 2799 & 768 & 17 & 24576 & 7 \\
21 & 2 & 1024 & 378 & 32768 & 7 \\
24 & 28 & 1344 & 2 & 98304 & 1 \\
32 & 3878 & 1536 & 24 & 172032 & 1 \\
48 & 38 & 2048 & 161 & 196608 & 1 \\
64 & 3412 & 2688 & 2 & 1376256 & 1 \\
\hline
\end{tabular}
\end{center}
\end{table}

The distance distributions of the $(12,256,3)$
codes are of the form
\[
\begin{array}{l}
(1,0,0,16+\mu,39-\mu,48-4\mu,48+4\mu,48+6\mu,\\
39-6\mu,16-4\mu,4\mu,\mu,1-\mu),
\end{array}
\]
where $0 \leq \mu \leq 1$ (the distance distribution is
unique for the other tabulated parameters). The distribution of the
value of $\mu$ amongst these codes is shown in Table~\ref{tab:dist}.

\begin{table}[htbp]
\begin{center}
\caption{Distance distributions of $(12,256,3)$ codes}
\label{tab:dist}
\begin{tabular}{rrrrrrrr}\hline
$256\mu$ & \# & $256\mu$ & \# & $256\mu$ & \# & $256\mu$ & \# \\ \hline
0 & 127 &  128 & 3719 & 172 & 184 &   216 & 7787 \\
32 & 132 & 132 & 15 &   176 & 2703 &  220 & 2298 \\
60 & 4 &   136 & 269 &  180 & 142 &   224 & 23319 \\
64 & 720 & 140 & 3 &    184 & 1424 &  228 & 2091 \\
84 & 6 &   144 & 403 &  188 & 313 &   232 & 9405 \\
88 & 37 &  148 & 35 &   192 & 17343 & 236 & 2253 \\
96 & 1055 &152 & 105 &  196 & 1003 &  240 & 11324 \\
108 & 18 & 156 & 133 &  200 & 2445 &  244 & 1746 \\
112 & 181 &160 & 5149 & 204 & 1112 &  248 & 3779 \\
116 & 24 & 164 & 47 &   208 & 11370 & 252 & 602 \\
124 & 6 &  168 & 209 &  212 & 1578 &  256 & 120992 \\\hline
\end{tabular}
\end{center}
\end{table}

It is known \cite{OP10} that not all $(12,256,3)$ and
$(13,512,3)$ codes can be lengthened to $(15,2048,3)$ codes
(and analogously for the extended codes
with $d=4$). In \cite{OP10} two equivalence classes of
$(13,512,3)$ codes that cannot be lengthened were found,
in addition to the 117819 equivalence classes that
can be lengthened. Our results show that the two
exceptional codes found in \cite{OP10} are the only
ones with this property. Moreover, they have equivalent
extensions, so there is a unique $(14,512,4)$ code
that cannot be lengthened to a $(16,2048,4)$ code; the
automorphism group of this code has order 768.

There are 10 equivalence classes of $(12,256,3)$ codes that
cannot be lengthened to $(15,2048,3)$ codes, and these have $3$ inequivalent
extensions. Codes from 7 of the 10 equivalence classes can be lengthened to
$(13,512,3)$ codes, which must then be equivalent to
the codes discovered in \cite{OP10}. The three equivalence
classes of $(12,256,3)$ codes that cannot be lengthened to
$(13,512,3)$ codes have equivalent extensions; the
unique $(13,256,4)$ code that cannot be lengthened to a
$(14,512,4)$ code has an automorphism group of order 384.

It turns out that one detail in \cite{OP10} is incorrect:
shortening the (two) $(13,512,3)$ codes that cannot be lengthened
to $(15,2048,3)$ codes always leads to $(12,256,3)$ codes
that cannot be lengthened to $(15,2048,3)$ codes.

Switching is a method for obtaining new codes from old ones.
See \cite{S01} for some general results on switching perfect
codes and \cite{OPP10} for specific results regarding
$(15,2048,3)$ perfect codes.

In \cite{OP10} it is shown that there are at least 21 switching
classes of $(13,512,3)$ codes. As no new $(13,512,3)$ codes were
discovered in the current classification, 21 is the exact number of
switching classes. The number of codes in the switching classes
is 115973, 1240, 561, 6 (2 classes), 4, 3 (6 classes), 2 (6 classes),
and 1 (3 classes).
The $(12,256,3)$ codes are partitioned into 10 switching classes
of the following sizes: 234749, 2509, 331, and
3 (7 classes).

The sets of codewords affected when switching are called
\mbox{$i$-\emph{components}.} Various information regarding $i$-components of
the $(15,2048,3)$ codes is provided in \cite{OPP10}. For the
$(12,256,3)$ and $(13,512,3)$ codes, the possible sizes of minimal
\mbox{$i$-components}
are 16, 32, 64, 96, 112, and 128; and
32, 64, 128, 192, 224, and 256, respectively.

Last but not least, the classification approach developed here
provides an alternative---and faster, starting from scratch---way
for classifying the $(15,1024,4)$ and $(16,2048,4)$ codes, which
was first done in \cite{OP09}.

\section{Lengthening $2^m-4$ and $2^m-3$ Codes}

\label{sect:lengthen}

The examples of $(12,256,3)$ and $(13,512,3)$ codes that cannot
be lengthened to $(15,2024,3)$ codes
lead to the obvious question
whether there---for some or all $m\geq 5$---are optimal codes of
length $2^m-4$ and $2^m-3$ that cannot be
lengthened to perfect
codes of length $2^m-1$. We shall now show that such codes indeed
exist for all such $m$. Before the construction, we consider a
necessary condition for a code to be a
triply-shortened perfect code; this question is studied in
greater depth in~\cite{K10a, K10b}.

The \emph{neighbors} of a word is the set of words at Hamming
distance 1. The \emph{complement} of a binary word is obtained by
adding the all-one vector to the word. Similarly, the complement
of a code $C$, denoted by $\overline C$, consists of the complements
of its codewords.

\begin{lemma}
\label{lem:complement}
Let\/ $C$ be an even\/ $(n=2^m-3,M=2^{2^m-m-4},4)$ code, and let\/
$E = \{{\bf x} \in \F_2^n : d({\bf x}, \overline C) \geq 3,\ \wt({\bf x})\ \mbox{even}\}$,
$\overline E = \{{\bf x} \in \F_2^n : d({\bf x},C) \geq 3,\ \wt({\bf x})\ \mbox{odd}\}$.
A word of\/ $\overline E$ has on average one neighbor in\/ $E$.
\end{lemma}
\begin{IEEEproof}
By Corollary~\ref{cor:weightdist},
$C$ has a unique distance distribution $A_i$, especially $A_{n-1} = 1$
and $A_{n-3} = (n-1)(n-5)/6$.

Since $A_{n-1} = 1$ and there cannot be more than one codeword at distance
$n-1$ from some codeword, it follows that each codeword of $C$ has exactly
one neighbor in $\overline C$. We define
\[
D = \{{\bf x} \in \F_2^n : d({\bf x}, \overline C) = 1\} \setminus C.
\]
Note that $|D| = (n-1)M$.

Let $E$ be the set of even words
in $\F_2^n$ that do not belong to $C \cup D$. The size of the set $E$ is
$2^{n-1}-|C|-|D| = (2^m - 1 - (2^m-4))M = 3M$. Similarly the odd-weight
words of $\F_2^n$ are divided into $\overline C$, $\overline D$, and $\overline E$.

We now define
\begin{equation*}
p(A, B) = \frac{1}{|A|}|\{({\bf a}, {\bf b}) : {\bf a} \in A, {\bf b} \in B,
d({\bf a}, {\bf b}) = 1\}|,
\end{equation*}
which gives the average number of neighbors in $B$ for a word in $A$.

Let us first count $p(D, \overline D)$.
For every pair ${\bf d} \in D$, ${\bf d'}\in \overline D$ with
$d({\bf d},{\bf d'})=1$, there are unique
${\bf c'} \in \overline C$, ${\bf c} \in C$ at distance $1$ from
${\bf d}$ and ${\bf d'}$, respectively; moreover, $d({\bf c},{\bf c'})$
is 1 or 3. For the case $d({\bf c},{\bf c'})=1$, there are
$MA_{n-1}$ possibilities to choose ${\bf c}$ and ${\bf c'}$,
each corresponding to $n-1$ pairs $({\bf d},{\bf d'})$. For the case
$d({\bf c},{\bf c'})=3$, there are
$MA_{n-3}$ possibilities to choose ${\bf c}$ and ${\bf c'}$,
each corresponding to 6 pairs $({\bf d},{\bf d'})$.
The total number of pairs $({\bf d},{\bf d'})$ is then
$P = (n-1)MA_{n-1}+6MA_{n-3}$, so
\[
p(D, \overline D) = \frac{P}{|D|} = \frac{M(n-1+6(n-1)(n-5)/6)}{M(n-1)}=n-4.
\]

Since $p(D, \overline C) = 1$ by the definition of $D$, we get that
$p(D, \overline E) = n - p( D,\overline C) - p( D, \overline D) = 3$,
$p(\overline E,D) = p(D,\overline E)|D|/|\overline E|=n-1$,
and $p(\overline E,E) = n - p(\overline E,D) = 1$.
\end{IEEEproof}

We define the \emph{conflict graph} of a code $C$ with minimum distance
$d$ as the graph with one vertex for each word that is at distance at
least $d-1$ from $C$ and with edges between vertices whose
corresponding words are at mutual distance less than $d$
(this is essentially the complement of a compatibility graph; see
Section~\ref{sect:result}). When we are specifically considering
\emph{even-distance} codes, we modify this
definition and only consider words that are at odd distance from $C$.

\begin{theorem}
\label{thm:tripartite_ext}
An\/ $(n=2^m-3, M=2^{2^m-m-4}, 4)$ code\/ $C$ is a triply-shortened extended perfect code
if and only if its conflict graph is tripartite, that is, is\/ $3$-colorable.
\end{theorem}
\begin{IEEEproof}
W.l.o.g., $C$ is an even code.
By the proof of Lemma~\ref{lem:complement}, the conflict graph of
$C$ has order $3M$.

Assume that $C$ is a triply-shortened
extended perfect code. As the extended perfect code is self-complementary,
it has the form
\[
\begin{array}{l}
C000 \cup D 001 \cup E 010 \cup F 100 \cup\\
\overline C 111 \cup \overline D 110 \cup \overline E 101 \cup \overline F 011,
\end{array}
\]
for some $(n, M, 4)$ codes $D$, $E$, and $F$ with odd weights.
Furthermore $D$, $E$, and $F$ must be independent sets in the conflict graph of $C$,
so the conflict graph is tripartite.

To prove implication in the opposite direction, we assume that
the conflict graph of the (even) code $C$ is tripartite
with parts $D$, $E$, and $F$. Now construct the code
\[
C00 \cup D01 \cup E10 \cup \overline F 11,
\]
which is an even code. Each of the four parts of this code has minimum
distance at least $4$. Moreover, from the definition of a conflict graph
and the fact that $D \cap E = \emptyset$,
$C00 \cup D01 \cup E10$ has minimum distance at least $4$.
For every word ${\bf c} \in C$, there is a word
${\bf c'} \in C$ such that $d({\bf c},{\bf c'})=n-1$, so
$\overline {\bf c} \not\in F$ (otherwise we would have
$d(C,F)=1$ which is not possible) and thereby
$\overline C \cap F = \emptyset$, which further
implies that $C00 \cup \overline F 11$ has
minimum distance at least $4$.

Since $D$, $E$, and $F$ have minimum distance at least 4 and
$|D|+|E|+|F|=3M$, where $M=2^{2^m-m-4}$, it follows that
$|D|=|E|=|F|=M$, and all of these codes are
optimal $(n=2^m-3, M=2^{2^m-m-4}, 4)$ code. Hence
every word in $F$ is at distance $n-1$ from exactly
one other word in $F$, whereby every word in $F$
has exactly one neighbor in $\overline F$. Using this result and
the fact, by Lemma~\ref{lem:complement}, that every word in $F$ has
on average one neighbor in $\overline D \cup \overline E \cup \overline F$,
we get that a word in $F$ has no neighbors in $\overline D \cup \overline E$.
Consequently, $d(D,\overline F)\geq 3$ and $d(E,\overline F)\geq 3$, so
$D01 \cup \overline F 11$ and $E10 \cup \overline F 11$ have minimum
distance at least 4.

Now we have lengthened $C$ to a $(2^m-1, 2^{2^m-m-2}, 4)$ code, which
has a (unique) lengthening to an extended perfect code \cite{B99}.
\end{IEEEproof}

\begin{corollary}
\label{lem:tripartite}
An\/ $(n=2^m-4, M=2^{2^m-m-4}, 3)$ code is a triply-shortened perfect code
if and only if its conflict graph is tripartite, that is, is\/ $3$-colorable.
\end{corollary}
\begin{IEEEproof}
Extend the code (to get even weights only) and the words in the conflict
graph (to get odd weights only), and use
Theorem~\ref{thm:tripartite_ext}.
\end{IEEEproof}

Now we proceed to the construction of codes that cannot be
lengthened to perfect codes.
We start with a lemma, which is followed by the main result of
this section.

\begin{lemma}
\label{lem:partition}
The space\/ $\F_2^{13}$ (resp.\ $\F_2^{12}$) can be partitioned into\/ $16$
copies of\/ $(13,512,3)$ codes
(resp.\ $(12,256,3)$ codes),
where at least one of the
codes cannot be lengthened to a\/ $(15,2048,3)$ code.
\end{lemma}

\begin{IEEEproof}
We construct a partition of $\F_2^{13}$, where one
of the codes is a $(13,512,3)$ code $C$ with a
$(12,256,3)$ subcode, neither of which can be extended
to a $(15,2048,3)$ code; such codes exist by \cite{OP10}
and Section~\ref{sect:res}. With the desired partition
for $\F_2^{13}$, shortening then provides a partition for
$\F_2^{12}$.

We know \cite{OP10} that $C$ can
be obtained by switching a code $C'$
that can be lengthened to some $(15,2048,3)$ code $D$.
Assume that $C'$ is obtained by shortening with respect to
the $0$s in the first two coordinates
of $D$ and that the switch with which $C$ is obtained from
$C'$ makes changes to the first coordinate of $C'$.

Via $D,D+{\bf e}_1,D+{\bf e}_2,\ldots$, we get a
partition of $\F_2^{15}$ into $16$ $(15,2048,3)$ codes.
By repeated shortening of these codes, one gets partitions
of $\F_2^{n}$ into $16$ $(n,2^{n-4},3)$ codes. If shortening
is carried out with respect to the $0$s in the first two
coordinates, then $C'$ is one of the $16$ codes $(13,512,3)$
codes that partition $\F_2^{13}$, and so is the
(equivalent) code $C'' = C'+{\bf e}_1$.

The fact that $C$ can be obtained from $C'$ by
changing only some values in the first coordinate of
$C'$ together with the observation that $C' \cup C''
= C \cup (C + {\bf e}_1)$ shows that $C'$ and $C''$
can be replaced in the partition of $\F_2^{13}$
by two codes neither of which can be lengthened to
a $(15,2048,3)$ code.
\end{IEEEproof}

\begin{theorem}
\label{thm:nonshortened}
For\/ $m \ge 4$, there are\/ $(2^{m}-4, 2^{2^m-m-4}, 3)$ codes and\/
$(2^{m}-3, 2^{2^m-m-3}, 3)$ codes that cannot be lengthened to a
perfect code of length\/ $2^m-1$.
\end{theorem}

\begin{IEEEproof}
We consider the case of length $2^m-4$.
Let $P$ be a perfect one-error-correcting code
of length $s = 2^{m-4}-1$, and
let $D_0, \ldots, D_{15}$ be the partition of $\F_2^{12}$ from
Lemma~\ref{lem:partition}, where $D_0$ can be lengthened to an
optimal code of length 13 but not to a perfect code of length 15.
Furthermore, let $A_0^0,\ldots, A_{15}^0$ be a
partition of the even-weight words of $\F_2^{16}$ into extended
perfect codes (for example, take cosets of the extended Hamming code),
and let $A_0^1,\ldots, A_{15}^1$ be such a partition of the odd-weight
words of $\F_2^{16}$.

Now consider the code
\begin{equation}
\label{eq:codec}
C = \bigcup_{\substack{\sum_{j=1}^{s+1}i_j \equiv 0 \!\!\!\!\pmod{16}\\
(x_1,\ldots ,x_s) \in P}}
A_{i_1}^{x_1} \times A_{i_2}^{x_2} \times \cdots \times A_{i_s}^{x_s} \times D_{i_{s+1}}
\end{equation}
of length $2^m-4$. It is not difficult to show that the code $C$,
the construction of which is a variation of a construction in \cite{P84},
has the desired minimum distance, length, and cardinality. Since the
conflict graph of $C$ contains as a subgraph the conflict graph of
$D_0$, which is not tripartite, the conflict graph of $C$ cannot be
tripartite either. It then follows from Corollary~\ref{lem:tripartite}
that $C$ cannot be lengthened to a perfect one-error-correcting code of
length $2^m-1$.

Since the partition $D_0, \ldots, D_{15}$ was chosen so that it can be
lengthened to a partition $D'_0, \ldots, D'_{15}$ of $\F_2^{13}$, the
code $C$ can be lengthened to a $(2^m-3, 2^{2^m-m-3}, 3)$ code that cannot
be lengthened further---alternatively,
use the partition $D'_0, \ldots, D'_{15}$ instead in (\ref{eq:codec}).
\end{IEEEproof}

\begin{corollary}
For\/ $m \ge 4$, there are\/ $(2^{m}-3, 2^{2^m-m-4},4)$ codes and\/
$(2^{m}-2, 2^{2^m-m-3},4)$ codes that cannot be lengthened to an
extended perfect code of length\/ $2^m$.
\end{corollary}

\begin{IEEEbiographynophoto}{Denis S. Krotov} 
was born in Novosibirsk, Russia in October 1974. He
received a Bachelor's degree in mathematics from Novosibirsk 
State University in 1995,
a Master's degree in 1997 from the same university, and a Ph.D.\
degree in mathematics from Sobolev Institute of Mathematics, Novosibirsk,
in 2000. Since 1997, he has been with Theoretical Cybernetics Department,
Sobolev Institute of Mathematics, where he is currently a Senior Researcher.
In 2003, he was a Visiting Researcher with Pohang University of Science and
Technology, Korea. His research interest includes subjects related to
combinatorics, coding theory, and graph theory.
\end{IEEEbiographynophoto}

\begin{IEEEbiographynophoto}{Patric R. J. \"Osterg{\aa}rd}
was born in Vaasa, Finland, in 1965. He received the M.Sc.\ (Tech.)\
degree in electrical engineering and the D.Sc.\ (Tech.)\ degree in
computer science and engineering, in 1990 and 1993, respectively,
both from Helsinki University of Technology TKK, Espoo, Finland.

From 1989 to 2001, he was with the Department of Computer Science
and Engineering at TKK. During 1995--1996, he visited Eindhoven
University of Technology, Eindhoven, The Netherlands, and in
2010 he visited Universit\"at Bayreuth, Germany. Since 2000,
he has been a Professor at TKK---which merged with two other universities
into the Aalto University in January 2010---currently in the Department of
Communications and Networking. He was the Head of the Communications
Laboratory at TKK in 2006--2007. He is the coauthor of
\emph{Classification Algorithms for Codes and Designs}
(Springer-Verlag, 2006), and, since 2006, co-Editor-in-Chief of the
\emph{Journal of Combinatorial Designs}. His research interests
include algorithms, coding theory, combinatorics, design theory, and
optimization.

Dr.\ \"Osterg{\aa}rd is a Fellow of the Institute of Combinatorics
and its Applications. He is a recipient of the 1996 Kirkman Medal.
\end{IEEEbiographynophoto}

\begin{IEEEbiographynophoto}{Olli Pottonen}
was born in Helsinki, Finland, in 1984. He received the M.Sc.\
(Tech.)\ degree in engineering physics and the D.Sc.\ (Tech.)\ degree in
information theory from Helsinki University of Technology TKK,
Espoo, Finland, in 2005 and 2009, respectively. He was with
the Finnish Defence Forces Technical Research Centre in 2009--2010
and with the Department of Information and Computer Science at
Aalto University School of Science in 2010.
In 2011 he is visiting Universitat Polit\`ecnica de Catalunya
in Spain. His research interests include coding theory, combinatorics, 
algorithm design, and discrete mathematics in general.

Dr.\ Pottonen was awarded the Dissertation of the year 2009 prize
at TKK.
\end{IEEEbiographynophoto}

\end{document}